\documentstyle[12pt]{article}
\def\applt{\;\; {\lower3pt\hbox{$
{\buildrel < \over {\scriptstyle \sim} }$}}\;\;}

\large
\begin{document}
\newcommand{\cl}{\centerline}
\newcommand{\beq}{\begin{equation}}
\newcommand{\eeq}{\end{equation}}
\newcommand{\beqa}{\begin{eqnarray}}
\newcommand{\eeqa}{\end{eqnarray}}
\def\d{{\rm d}}
\hfill{ANL-HEP-PR-94-19}\par
\hfill{CCUTH-94-03}\par
\vskip 0.3 true in
\centerline{\large \bf The Transition to Perturbative QCD}
\centerline{\large \bf in Compton Scattering}
\vskip .4in
\centerline{Claudio Corian\`{o}\footnote{Work supported by
the U.S. Department of Energy, Division of
High Energy Physics, Contract W-31-109-ENG-38}
and Hsiang-nan Li$^2$}
\vskip 0.4in
\centerline{ ${^1}$High Energy Physics Division,
Argonne National Labobratory,}
\centerline{9700 South Cass, IL 60439, USA}
\vskip 0.3cm
\centerline{$^2$Department of Physics, National Chung-Cheng University,}
\centerline{Chia-Yi, Taiwan, R.O.C.}
\vbox {\vskip 3.0cm}
\centerline{ABSTRACT}
\vskip 0.5cm
We investigate the transition to perturbative
QCD in Compton processes, by concentrating on the specific reactions
$\pi \gamma \to \pi \gamma$ and pion photoproduction at moderate energy
scales. New sum rules for each of the helicities
involved in the scattering and the cross section are given,
together with a detailed stability analysis.
Our results are then compared with those derived from higher-power
factorization
incorporating Sudakov suppression for quark elastic scattering.
An overlap of predictions obtained by the two approaches is observed
at a momentum transfer as low as 4 GeV$^2$ and at a
photon scattering angle around 40$^o$.
Our work shows that
factorization theorems and sum rule methods complement each other
in the description of Compton scattering
at moderate energy scales, and give information
on the transition to perturbative QCD in these processes. The sum rule
formalism is further  applied to the
crossed reaction $\gamma \gamma\to \pi^+ \pi^-$, and is in very good
agreement the experimental data.

\newpage
\section{Introduction}
Exclusive processes have been one of the most challenging testing grounds
for perturbative QCD (PQCD).
Although the description of exclusive processes using PQCD is considered
successful for momentum transfer $Q^2$ going to infinity,
it remains a controversy whether PQCD is applicable to exclusive
processes at moderate  $Q^2$.
This issue has been widely discussed in the literature, however,
general agreement on it is still missing.
It has been shown that the standard leading-order and leading-twist
factorization \cite{BL} fails to give reliable
predictions to hadron form factors at experimentally accessible $Q^2$
because of the dominance of soft contributions from the end-points of
parton momentum fractions \cite{IL}.
More recently, the understanding of exclusive
processes has been improved by studying the transverse momentum
dependence of scattering amplitudes \cite{BS}, which is introduced by
transverse momenta carried by the
collinear partons that enter hard scatterings.
Modified perturbative expressions for pion and proton electromagnetic form
factors including this higher-power dependence
have been given \cite{LS,HN}.
It is found that the all-order summation of the higher-power effects
leads to suppression of end-point contributions,
and extends the applicability of PQCD down to 2-3 GeV. That is,
hadron form factors transit from non-perturbative to perturbative QCD
around this scale.

However, it is still unclear how far low in $Q^2$  the modified
perturbative picture is yet successful
for other more complicated exclusive processes.
In this paper we shall investigate the transition to PQCD in Compton
processes by comparing predictions from the above
modified perturbative formalism with those
from QCD sum rules \cite{SVZ},
a method to study non-perturbative properties of
hadrons. A sum rule approach to a simple Compton-type process,
pion Compton scattering, has been
discussed in refs. \cite{CRS} and \cite{co}, in which
the similarity between fixed-angle Compton scattering and form factors
was explored. Sum rules for the sum $H$ of the two helicity invariant
amplitudes, $H_1$ and $H_2$, involved in pion Compton scattering
was given in \cite{co}, and preliminarily
compared to the modified perturbative predictions in \cite{coli1}.
A transition scale at $Q^2$ as low as 4 GeV$^2$, for photon scattering
angle around 40$^o$, has been observed.
Compared to the sum rule formalism for form factors,
the analyticity region of
Compon scattering is more severely constrained \cite{CRS,co,co1}.
In order to maintain the stability of the sum rule for $H$ in the finite
analyticity region, we have proposed a modified phenomenological
parametrization for resonance and continuum states in \cite{coli}.

PQCD and sum rule approaches employ very different physical pictures:
the former takes a hard-scattering view of exclusive processes,
while the latter is described by the dominance of
``Feynman mechanism" \cite{NR}.
It has been pointed out \cite{LS,CRS}
that the PQCD approach should be complemented by sum rule methods
in order to fully understand the behaviors of simple elastic
scatterings at moderate energy scales.
The parallel discussion based on
both PQCD and sum rules is of interest, since it can help
to better understand the transition to PQCD in exclusive processes,
and clarify, to some extent, the questioned applicability of PQCD.

In this paper we shall investigate individual sum rules, for $H_1$ and $H_2$
respectively, in great detail, together with their modified stability
analysis \cite{coli}. The cross section of pion Compton scattering
is also given. All the steps are then repeated
using the modified PQCD formalism,
and predictions from the two approaches are compared.
We find that
perturbative contributions dominate at large angles,
while at smaller angles sum rule contributions are important.
The overlap between the two descriptions appears at about $40^0$
of photon scattering angle and at $Q^2$ around 4 GeV$^2$,
similar to the behavior observed in the sum rule for
$H$ \cite{coli1}. We then confirm the conclusion drawn
in \cite{FSZ} that large-angle pion Compton scattering can be treated
by pQCD reliably. To justify our formalism, the sum rule methods are
further applied to the crossed version of Compton scattering, pion
photoproduction $\gamma\gamma\to\pi^+\pi^-$, for which experimental
data are avaliable \cite{B}. It is found that our predictions are
consistent with the data.

This paper is organized as follows.
A brief illustration of the sum rule formalism is given in section 2,
where the main features of the
sum rule for $H$, extensively studied in \cite{co,coli1}, are reviewed.
We present in section 3 the complete
evaluation of the individual sum rules for $H_1$ and $H_2$, which
include both the local duality contribution
and the power corrections. The stability
analysis of the new sum rules based on the modified phenomenological
model \cite{coli} is performed. $H$ and $H_1$ are evaluated using the
PQCD formalism in section 4.
Section 5 contains numerical results of the two
helicities and the cross section derived by the two approaches, and we
specify the transition region.
The sum rule for pion photoproduction
is also analyzed, and the results are compared with the data.
Section 6 is the conclusion. Three appendices are inserted to
clarify the conventions
and to decribe some details of the calculation.

\section{Sum Rules for $H$}

As the relevant energy scales approach a
resonance region, a new complexity in the description of exclusive
processes shows
up, because non-perturbative effects, related
to the dynamical behaviors of QCD vacuum,  can no longer be neglected.
Such effects  can not be incorporated into a direct
perturbative treatment based on factorization theorems, and
to keep them into account
one must resort to completely different methods, the most successful
one being QCD sum rules.
Sum rule methods have been recently extended to fixed-angle Compton
scattering from their usual application to two- and three-point processes
in \cite{CRS}.
So far, all previous works have concentrated on the specific
combination of the invariant amplitudes, $H=H_1+H_2$.
We have communicated before in a brief letter \cite{coli1}
that sum rule predictions for $H$ dominate
at intermediate photon scattering angles over perturbative calculations.

QCD sum rules connect the timelike region of a suitably chosen correlator
to the spacelike region in the complex planes of external virtualities
$p_i^2$ by a dispersion relation. For pion Compton scattering the relevant
correlator corresponds to the lowest-order diagrams without virtual
gluons, as shown in fig.~1.
In the case of four-point functions the
existence of an analiticity region for the correlator in the $p_i^2$ planes,
at fixed Mandelstam invariants $s$, $t$ and $u$, is far from being
obvious.

In the deep Eucliden region, the spacelike region with
large negative virtualities, the operator product expansion (OPE)
for the correlator is usually valid, and can be calculated
using perturbation theory.
The result is organized in terms of a lowest-order perturbative contribution
plus power corrections,
the latter being parametrized by the lowest dimensional QCD condensates.
These corrections can be obtained (to lowest
order in $\alpha_s$) in two standard ways: by Cutkosky rules,
and by Borel transforms \cite{co1}.
Both methods \cite{IS} \cite{NR} have been extensively employed
for three-point functions. The regions of analyticity in the case of massive
and massless correlators have been described in detail in ref.~\cite{co1}.
Radiative corrections to the dispersion relations, which enter in the
description of these processes, are also calculable by the methods
discussed in ref.~\cite{co1}.

The timelike region of the correlator cannot be described by perturbation
theory for the virtualities close to the lowest resonances,
and the residues at the resonance poles are
known only up to their symmetry properties. Therefore, a phenomenological
model parametrizing the resonance and the continuum contributions has to be
proposed. The continuum contribution is usually chosen as the perturbative
part on the Operator Product Expansion side of the sum rules.

For forward or backward scattering the Compton processes
cannot be described by the extended sum rule formalism.
A simple observation of this fact
is that the involved leading spectral function becomes singular
as the Mandelstam invariants $s$ and $t$ vanish \cite{co1}. Hence,
the OPE is applicable only in a suitably chosen angular range.
In order not to include
extra $u$-channel singularities, we study the correlator in
a finite region in the $p_i^2$ planes.
Even for a finite analiticity region, the existence of a
dispersion relation guarantees that the behavior
in the timelike region can be related to that in the spacelike region.
This relies
on the introduction of a modified Borel transform \cite{CRS},
with finite radius $\lambda^2$, to characterize the
analyticity region.
A spurious dependence on the "Borel radius" $\lambda^2$
is then brought into the dispersive representation
of the coefficients of such corrections. As discussed in \cite{coli},
this $\lambda$-dependence is unphysical and spoils the stability of the
sum rule.

The coefficients of the OPE are usual Feynman integrals,
now in a dispersive form \cite{co}.
It is possible, however, to extend the dispersive representation of
each Feynman integral (and also of the power corrections) to all
positive values of $s_1$ and $s_2$, in the case of massless correlators.
For pion photoproduction the pion pole is approximately set at
$p_1^2=p_2^2=0$. If one wants to extract the contribution of additional
states from the phenomenological side of the sum rule or, for instance,
extend the method suggested in ref.~\cite{CRS} to the proton case,
then the spectral functions evaluated by the OPE
have an explicit mass dependence and additional singularities compared to
the massless case. These singularities, specifically, are described by
Landau surfaces and disappear as the mass in the correlator is set to be zero.
The discussions of all these issues is in ref.~\cite{co1}.

{}From a physical viewpoint, it is expected that at large
virtualities
the spectral densities should give negligible contribution to
the dispersion integral. It is also easy to show that most of the
contribution to the spectral density comes from the region of approximately
equal virtualities $s_1$ and $s_2$ ($s_1\approx s_2$).

Based on these arguments, we have proposed a
modified phenomenological model for resonance and continuum states
in \cite{coli} to avoid the undesirable $\lambda^2$ dependence.
This new ans\"{a}zte removes the contributions of large
virtualities both from the phenomenological side
and from the OPE side of the sum rules.
In other words, according to this model for the continuum
(on what is termed "the phenomenological side" of the sum rule),
all the coefficients of the OPE cannot contribute to the sum rule for
large virtualities.

Application of this ans\"{a}zte to the pion form factor \cite{coli}
gives results which
are not substantially different from what have been obtained in the
literature \cite{NR,IS}.
A stability analysis of the sum rule for $H$
based on the modified phenomenological model has been performed
in \cite{coli}, where
the local duality interval $s_0$ and the Borel mass $M$ are found to
take values very
close to those in the form factor case \cite{NR,IS}.
In order to make our discussion self-contained, we briefly
summarize the derivation of the
sum rule for $H$ \cite{CRS}. We then compute
$H_1$ and $H_2$, from which the
cross section is derived accordingly.

We start with the following four-point correlator
\begin{eqnarray}
\Gamma_{\sigma\mu\nu\lambda}(p_1^2,p_2^2,s,t)&=&i\int{\rm d}^4x {\rm d}^4y
 {\rm d}^4z\exp
(-ip_1\cdot x+ip_2\cdot y-iq_1\cdot z)
\nonumber \\
& &\times \langle 0|T\left(\eta_{\sigma}(y)J_{\mu}(z)J_{\nu}(0)
\eta_{\lambda}^{\dagger}(x)\right)|0\rangle \; ,
\label{tp}
\end{eqnarray}
where
\begin{eqnarray}
J_{\mu}=\frac{2}{3}\bar{u}\gamma_{\mu}u -\frac{1}{3}\bar{d}\gamma_{\mu}d,
\;\;\;\;\;\;
\eta_{\sigma}=
\bar{u}\gamma_5\gamma_{\alpha}d
\label{jd}
\end{eqnarray}
are the electromagnetic and axial currents, respectively,
of $up$ and $down$ quarks.
The on-shell momenta $q_1$ and $q_2$ are carried by the two physically
polarized photons. The two pion momenta are denoted by $p_1$ and $p_2$,
with $s_1=p_1^2$ and $s_2=p_2^2$ the virtualities. The Mandelstam
invariants $s=(p_1 +q_1)^2$, $t=(p_2-p_1)^2$ and $u=(p_2-q_1)^2$
obey the relation $s + t +u\,=\,s_1 + s_2$.
The invariant amplitudes $H_1$ and $H_2$
involved in pion Compton scattering are then obtained through the expansion
of the matrix element
\beq
M_{\mu\nu}= i\int d^4z   e^{-iq_1\cdot z}
  \langle p_2|T\left(J_{\mu}(z)J_{\nu}(0)\right) |p_1 \rangle
  \label{mnula}
\eeq
for a specific time-ordering as
\beq
M_{\mu\nu}= H_1(s,t) e^{(1)}_{\mu}e^{(1)}_{\nu} +
H_2(s,t) e^{(2)}_{\mu}e^{(2)}_{\nu}\;.
\label{h1h2}
\eeq
The helicity vectors $e^{(1)}$ and $e^{(2)}$ are
defined in \cite{CRS,coli}, satisfying the orthogonality condition
$e^{(i)}\cdot e^{(j)}=-\delta_{ij}$.
The expression of the sum rule for $H$ is derived by contracting
$\Gamma_{\sigma\mu\nu\lambda}$ with $-g^{\mu\nu}n^{\sigma}n^{\lambda}$,
where
\beq
n^\mu = \biggl (e^{(2)} \pm ie^{(1)} \biggr )^\mu
\label{nconstruct}
\eeq
is a suitable projector in analogy to that for pion form factor \cite{NR}.
The phenomenological model employed in sum rules
are characterized by a resonance contribution from the double poles of
the pion
states, and by a continuum one $\Delta^{\rm pert}$ \cite{CRS,co}
for $p_i^2 > s_0$, which
is exactly the perturbative contribution on the OPE side:
\beqa
\Delta(p_i^2,s,t)&=&
f_\pi^2  n\cdot p_{1} n\cdot p_{2}  (2\pi)^2\delta(p_1^2-m_\pi^2)
\delta(p_2^2-m_\pi^2) \nonumber \\
&  & \times H(s,t)+ \Delta^{\rm pert}(p_i^2,s,t)
[1-\theta(s_0-p_1^2)\theta(s_0-p_2^2)]\;.
\label{oml}
\eeqa

A modified Borel transform \cite{CRS}
\begin{equation}
{\cal B}' = \oint_{C}\frac{dp_1^2}{M_1^2}\oint_{C}\frac{dp_2^2}{M_2^2}
e^{-p_1^2/M_1^2}e^{-p_2^2/M_2^2}
\left (1 - e^{-(\lambda^2-p_1^2)/M_1^2} \right ) \left (1 - e^{-(\lambda^2-
p_2^2)/M_2^2} \right)
\label{bo}
\end{equation}
was then proposed to define the analyticity region,
with $C$ a contour of finite radius $\lambda^2$,
which is allowed to vary from $s_0$ to $(s +t)/2$.
The factor $1-\exp[-(\lambda^2-p_i^2)/M^2]$ in (\ref{bo}),
which is new compared to the standard Borel transform
\cite{NR,IS}, gives the constraint $p_i^2 < \lambda^2$,
and thus excludes the $u$-channel resonances from the spectral density,
ensuring that the transform is still regular
when $C$ crosses the branch cuts in the $p_i^2$ planes.
This modification introduces the extra unphysical parameter $\lambda^2$,
in addition to the usual Borel mass $M^2$.
Hence, the stability of the sum rule has to be established in the
two-dimensional $M^2$-$\lambda^2$ plane.
Using eq.~(\ref{oml}), the sum rule for $H$ has been
derived \cite{CRS}, in which the argument $s_i$ of
the perturbative spectral density on the OPE side runs up to $s_0$,
due to the cancellation by the continuum contribution on the
phenomenological side; while $s_i$ in the power corrections runs up to
$\lambda^2$.
A naive numerical analysis based on the above formalism has been
performed, and results
show a strong sensitivity to the variation of $\lambda^2$ \cite{coli}.

The strong $\lambda$ dependence of the sum rule is not desired, since
$\lambda$ is an unphysical parameter, and a physical quantity
like the invariant amplitude should be insensitive to it.
Therefore, we took an alternative approach to avoid the $u$-channel
singularities. We have proposed a modified phenomenological model
with the continuum contribution $\Delta^{\rm pert}$ replaced by
$\Delta^{\rm OPE}$, the full spectral expression
on the OPE side of the sum rule.
This modification makes sense, because the region with large virtualities
$p_i^2 > s_0$ can be regarded as perturbative, and an OPE
is allowed. With this choice we are requiring that
the phenomenological parametrization for the continuum
from $s_0$ to $\lambda^2$ truncates not only
the perturbative part, but also
the power corrections, on the OPE side.
Since the contributions from the large-virtuality
region $(s_0, \lambda^2)$ have been removed
by the above cancellation, $s_i$'s  never reach the upper bound $\lambda^2$,
and the remaining $\lambda$ dependent factors
$1- \exp[-(\lambda^2-p_i^2)/M^2]$ from the transform (\ref{bo})
can be dropped. Therefore, the overall
dependence on the Borel radius
disappears completely from the sum rule. In conclusion, we suggested
a modified phenomenological model with the original Borel transform
\begin{equation}
{\cal B} = \oint_{C}\frac{dp_1^2}{M_1^2}\oint_{C}\frac{dp_2^2}{M_2^2}
e^{-p_1^2/M_1^2}e^{-p_2^2/M_2^2}\;,
\label{obo}
\end{equation}
instead of a new transform with the original model as in
\cite{CRS}.
The resulting asymptotic expression for $H$ based on the modified
model is written as
\beqa
&&{f_\pi}^2H(s,t)\left({s(s+t)\over -t}\right)=\nonumber \\
& &\hspace*{0.5cm}\left(\int_{0}^{s_0}ds_1\int_{0}^{s_0}ds_2
\rho^{\rm pert}+\frac{\alpha_s}{\pi}\langle G^2 \rangle
\int_{0}^{s_0}ds_1\int_{0}^{s_0}ds_2\rho^{\rm gluon}\right)
e^{-(s_1+s_2)/M^2} \nonumber \\
& &\hspace*{0.5cm}+C^{\rm quark}\pi\alpha_s
\langle (\bar{\psi}\psi)^2 \rangle
\label{s_2}
\eeqa
for large invariant $Q^2$ \cite{co},
\beqa
Q^2={1\over 4}\left(s_1 + s_2 -t + \sqrt{(s_1 +s_2 -t )^2 - 4 s_1 s_2}
\right)\;,
\label{q}
\eeqa
where the perturbative, gluonic and quark contributions are
given by, respectively,
\beqa
\rho^{\rm pert}&=&{ 2560 Q^{14} \tau^{\rm pert}(s,Q^2,s_1,s_2)\over
3\pi^2  (s-2 Q^2 )(4 Q^4- s_1 s_2)^5 (s_1 s_2-2 Q^2 s)}\;,\nonumber \\
\rho^{\rm gluon}&=&{ 20480 Q^{22}(s - 2 Q^2)
\tau^{\rm gluon}(s,Q^2,s_1,s_2)\over
27 s (2 Q^2 -s_1)^2  (2 Q^2 -s_2)^2 (4 Q^4 - s_1 s_2)^5 (2 Q^2 s - s_1 s_2)
(4 Q^4 -2 Q^2 s + s_1 s_2)^2}\;, \nonumber \\
C^{\rm quark}&=&-{16\over 9}
{(8 M^4 s^2 +8 M^4 s t +8 M^2 s^2 t +8 M^2 s t^2 +4 s^2 t^2
+4 s t^3+ t^4)\over M^4 t^3}\;,
\label{oc}
\eeqa
with
\beqa
\tau^{\rm pert}&=&(s- Q^2)^2
(2 Q^4 s s_1 - Q^2 s^2 s_1 + 2 Q^4 s s_2 - Q^2 s^2 s_2 \nonumber \\
&&- 2 Q^4 s_1 s_2 - 6 Q^2 s s_1 s_2 + 3 s^2 s_1 s_2)\;, \nonumber \\
\tau^{\rm gluon}&=&-8 Q^{12} s -8 Q^{10} s^2 + 68 Q^8 s^3 -64 Q^6 s^4+
16 Q^4 s^5 \nonumber \\
& &+8 Q^{10} s s_1 +8 Q^8 s^2 s_1 -108 Q^6 s^3 s_1 +104 Q^4 s^4 s_1
-26 Q^2 s^5 s_1 \nonumber \\
& &+8 Q^{10} s s_2 +8 Q^8 s^2 s_2 -108 Q^6 s^3 s_2
+ 104 Q^4 s^4 s_2 -26 Q^2 s^5 s_2\;. \nonumber \\
& &
\label{gluonic}
\eeqa
Approximate methods for the evaluation of power corrections have been
developed in ref.~\cite{co}.
A further analysis of the diagrammatic expansion \cite{co1}
shows that it is possible to take the limit of
$\lambda \to  \infty$ in the dispersive
representation of the coefficients of the power corrections,
since their the spectral densities
are globally well defined
for massless correlators. The rigorous proof of these statements is
discussed in ref.~\cite{co1}.
Note the upper bound $s_0$ instead of $\lambda^2$
in the integral for the gluonic power correction
due to the cancellation from the phenomenological side.
The gluon and quark condensates,
$\langle G^2\rangle$ and $\langle (\bar{\psi}\psi)^2\rangle$,
take the values
\begin{eqnarray}
&&\frac{\alpha_s}{\pi}\langle G^2\rangle=1.2\times 10^{-2}{\rm GeV}^4\nonumber
\\
&&\alpha_s\langle (\bar{\psi}\psi)^2\rangle=1.8\times 10^{-4}{\rm GeV}^6\;.
\label{vev}
\end{eqnarray}

Since the $\lambda$ dependence is removed completely,
the stability analysis of (\ref{s_2}) is straightforward
following a method similar to \cite{NR,IS}.
We concentrated simply on the variation of $H$ with
respect to $M^2$.
It has been found that as $s_0=0.6$ GeV$^2$
there is the largest $M^2$ interval,
in which $H$ is approximately constant. Therefore, $s_0=0.6$ GeV$^2$
is the best choice which makes both sides of the sum rule most coincident.
This value of the duality interval is close to that given
in the form factor case, and consistent with its conjectured value of
0.7 GeV$^2$ in \cite{coli1}.
Different sets of $s$ and $t$ have been investigated. The best value of
$s_0$ does not vary significantly, and $H$ is almost constant
within the range $2 < M^2 < 6$ GeV$^2$.

Results for $H$ at different photon scattering angles $\theta^*$
in the Breit frame,
$\sin(\theta^*/2)=-t/(s-u)$, with $s_0=0.6$ and $M^2=4$
GeV$^2$ have been obtained \cite{coli1,coli}.
Basically, they show a similar dependence on
angles and momentum transfers $|t|$ to those derived using local duality
approximation \cite{coli1}.
These predictions have been
compared to the perturbative predictions obtained from the modified
factorization formula \cite{coli1}, which will be discussed in detail below.
Sum rule results are always larger than the
perturbative results at smaller angles.
The transition to PQCD appears at
about $|t|=4$ GeV$^2$ and at $\theta^*=40^o$,
where the perturbative contributions begin to dominate.

\section{Sum Rules for $H_1$ and $H_2$}

Following the formalism outlined in section 2, we derive sum rules
for the helicity invariant amplitudes $H_1$ and $H_2$,
which are extracted by
contracting $\Gamma_{\sigma\mu\nu\lambda}$ with
$e^{(1)\mu}e^{(1)\nu}n^{\sigma}n^{\lambda}$ and
$e^{(2)\mu}e^{(2)\nu}n^{\sigma}n^{\lambda}$ respectively.
Similarly, the modified
phenomenological model for $H_i$ is given by
\beqa
\Delta_i(p_i^2,s,t)&=&
f_\pi^2  n\cdot p_{1} n\cdot p_{2}  (2\pi)^2\delta(p_1^2-m_\pi^2)
\delta(p_2^2-m_\pi^2) \nonumber \\
&  & \times H_i(s,t)+\Delta_i^{\rm OPE}(p_i^2,s,t)
[1-\theta(s_0-p_1^2)\theta(s_0-p_2^2)]\;.
\eeqa
The method to evaluate the perturbative spectral
density and the power corrections from quark and gluon condensates
is described in \cite{co}. The result can be organized in the form

\beqa
&&{f_\pi}^2H_i(s,t)\left({s(s+t)\over -t}\right)=\nonumber \\
& &\hspace*{0.5cm}\left(\int_{0}^{s_0}ds_1\int_{0}^{s_0}ds_2
{\rho_i}^{\rm pert}+\frac{\alpha_s}{\pi}\langle G^2 \rangle
\int_{0}^{s_0}ds_1\int_{0}^{s_0}ds_2\rho_i^{\rm gluon}\right)
e^{-(s_1+s_2)/M^2} \nonumber \\
& &\hspace*{0.5cm}+C_i^{\rm quark}\pi\alpha_s
\langle (\bar{\psi}\psi)^2 \rangle\;.
\label{h2}
\eeqa
The perturbative and gluonic contributions for $H_1$ are
given respectively by
\begin{eqnarray}
\rho_1^{\rm pert}&=&{-1280 Q^{14} \tau_1^{\rm pert}(s,Q^2,s_1,s_2)\over
3\pi^2  (s-2 Q^2 )(4 Q^4- s_1 s_2)^5 (s_1 s_2-2 Q^2 s)}\;,
\nonumber\\
\rho_1^{\rm gluon}&=&{ 20480 Q^{22}(s - 2 Q^2)
\tau_1^{\rm gluon}(s,Q^2,s_1,s_2)\over
27 s (2 Q^2 -s_1)^2  (2 Q^2 -s_2)^2 (4 Q^4 - s_1 s_2)^5 (2 Q^2 s - s_1 s_2)
(4 Q^4 -2 Q^2 s + s_1 s_2)^2}\;,
\nonumber \\
& &
\end{eqnarray}
with
\begin{eqnarray}
\tau_1^{\rm pert}&=&(s- Q^2)^2
(2 Q^4 s s_1 - Q^2 s^2 s_1 + 2 Q^4 s s_2 - Q^2 s^2 s_2) \nonumber \\
&&-(2Q^6s-13 Q^4s^2+12 Q^2 s^3-3 s^4) s_1 s_2\;,
\nonumber \\
\tau_1^{\rm gluon}&=& -36 Q^{10} s^2 -14 Q^8 s^3 +32 Q^6 s^4-
8 Q^4 s^5 \nonumber \\
& &+22 Q^{10} s s_1 +2 Q^8 s^2 s_1 +99 Q^6 s^3 s_1 -100 Q^4 s^4 s_1
+25 Q^2 s^5 s_1 \nonumber \\
& &+22 Q^{10} s s_2 +2 Q^8 s^2 s_2 +99 Q^6 s^3 s_2
-100 Q^4 s^4 s_2 +25 Q^2 s^5 s_2\;. \nonumber \\
& &
\end{eqnarray}
The quark power correction for $H_1$ is computed as
\beqa
\Delta_1^{\rm quark}&=&{{256 {Q^6} \left( 2 {Q^2} - s \right)
    \tau_1^{\rm quark }\over
   {{{\left( 2 {Q^2} - {p_1}^2 \right) }^2} {{{p_1}}^6}
     {{\left( 2 {Q^2} - {p_2}^2 \right) }^2} {{{p_2}}^6}
     \left( 4 {Q^4} - 2 {Q^2} s + {p_1}^2 {p_2}^2 \right) }}}\;,
\nonumber \\
\label{q1}
\eeqa
where
\beqa
&& \tau_1^{\rm quark}=
 \left( 4 {Q^6} - 8 {Q^4} s + 4 {Q^2} {s^2} - 4 {Q^4} {p_1}^2 +
       6 {Q^2} s {p_1}^2 - 3 {s^2} {p_1}^2 - 4 {Q^4} {p_2}^2\right.
\nonumber \\
&&\left.\,\,\,\,\,\,\,\,\,\,\,\,\,\, +
       6 {Q^2} s {p_2}^2 - 3 {s^2} {p_2}^2 \right)
     \left( {{{p_1}}^4} + {{{p_2}}^4} \right). \nonumber \\
\eeqa
{}From eq.~(\ref{q1}) $C_1^{\rm quark}$ is obtained through the application of
the standard Borel transform in eq.~(\ref{obo}) as
\begin{eqnarray}
C_1^{\rm quark}&=&{\cal B}(p_1^2\to M_1^2){\cal B}(p_2^2 \to M_2^2)
\Delta_1^{\rm quark}|_{M_1,M_2=M}\nonumber \\
&=&-{16\over 9}
{(8 M^4 s^2 +8 M^4 s t +16 M^2 s^2 t +16 M^2 s t^2 +4 s^2 t^2
+4 s t^3+ t^4)\over M^4 t^3}\;.
\end{eqnarray}

The corresponding quantities associated with $H_2$ are given by
\begin{eqnarray}
\rho_2^{\rm pert}&=&{1280 Q^{14} \tau_2^{\rm pert}(s,Q^2,s_1,s_2)\over
3\pi^2  (s-2 Q^2 )(4 Q^4- s_1 s_2)^5 (s_1 s_2-2 Q^2 s)}\;,
\nonumber\\
\rho_2^{\rm gluon}&=&{ 20480 Q^{22}(s - 2 Q^2)
\tau_2^{\rm gluon}(s,Q^2,s_1,s_2)\over
27 s (2 Q^2 -s_1)^2  (2 Q^2 -s_2)^2 (4 Q^4 - s_1 s_2)^5 (2 Q^2 s - s_1 s_2)
(4 Q^4 -2 Q^2 s + s_1 s_2)^2}\;,
\nonumber \\
C_2^{\rm quark}&=&\frac{128}{9}\frac{s(s+t)}{M^2t^2}\;,
\end{eqnarray}
with
\begin{eqnarray}
\tau_2^{\rm pert}&=&3(s- Q^2)^2
(2 Q^4 s s_1 - Q^2 s^2 s_1 + 2 Q^4 s s_2 - Q^2 s^2 s_2) \nonumber \\
&&-(4 Q^8+6Q^6s-39 Q^4s^2+36 Q^2 s^3-9 s^4) s_1 s_2\;,
\nonumber \\
\tau_2^{\rm gluon}&=& -8Q^{12}s+28 Q^{10} s^2 +82 Q^8 s^3 -96 Q^6 s^4+
24 Q^4 s^5 \nonumber \\
& &-14 Q^{10} s s_1 +6 Q^8 s^2 s_1 -207 Q^6 s^3 s_1 +204 Q^4 s^4 s_1
-51 Q^2 s^5 s_1 \nonumber \\
& &-14 Q^{10} s s_2 +6 Q^8 s^2 s_2 -207 Q^6 s^3 s_2
+204 Q^4 s^4 s_2 -51 Q^2 s^5 s_2\;. \nonumber \\
& &
\end{eqnarray}
It can be easily checked that the combination of the above two sum rules,
$H_1+H_2$, is equal to that of $H$ as shown in section 2.

A similar stability analysis of eq.~(\ref{h2}) is performed, and the
variation of $H_i$ with respect to $M^2$ for $s_0=0.5$-0.7 GeV$^2$
at $s=20$ and $|t|=10$ GeV$^2$ is shown in fig.~2.
The region on the right-hand side of vertical bars is the one where
the power corrections do not exceed 50\% of the perturbative contribution.
Obviously, in both of fig.~2a and 2b
the curve corresponding to $s_0=0.6$ GeV$^2$ exhibits a
largest $M^2$ interval, in which $H_i$ remains almost constant.
By varying $s$ and $t$, we find that the best choice for the duality
interval is $s_0=0.6$ GeV$^2$ and the corresponding stability region
is $2<M^2<6$ GeV$^2$, the same as those for $H$ \cite{coli}.
We shall evaluate $H_i$ by substituting $s_0=0.6$ and $M^2=4$ GeV$^2$
into eq.~(\ref{h2}) in section 5 in order to compare it with predictions
derived from the modified perturbative expression.
The detailed calculation of
all the relevant quantities in the sum rule approach is given in Appendix A.
The full expressions for the perturbative spectral functions of $H_1$ and
$H_2$ are exhibited in Appendix B.

\vskip 2.0cm

\section{The PQCD approach}

The modified pQCD formalism including transverse momentum dependence
of the hard scattering, which will be presented below, has been discussed
extensively in refs.~\cite{LS,HN,coli1}. We shall summarize the basic
reasoning leading to the modified perurbative expression, and review the
calculation of the invariant amplitude $H$ briefly \cite{coli}.
We then derive the first helicity amplitude $H_1$
using the same method. The second invariant amplitude is
obtained by $H_2=H-H_1$.

We consider the factorization formula for $H_1$
in the conjugate configuration, or $b$, space \cite{BS}
due to the introduction of transverse momenta,
$b$ being the separation between two valence quarks. The perturbative
expression includes the infinite summation of the higher-power effects
at lower momentum transfers, which
suppresses non-perturbative contributions.
This effect, termed ``Sudakov suppression",
exhibits a falloff of the invariant amplitudes at fixed $b$ faster
than any power of $Q$. The modified perturbative expression
reduces to the standard one as predicted by the
dimensional counting rules \cite{BL} at very large $Q^2$.
The higher-power effects appear as large logarithms resulting from
radiative corrections to the process, which will be analyzed below.

Basic diagrams for pion Compton scattering in the PQCD approach are shown
in fig.~3,
which differ from fig.~1 by an extra exchanged gluon.
In most cases, large logarithms from radiative corrections in exclusive
processes do not cancel, and have to be summed up.
The radiative corrections to fig.~3 can be classified
into two categories: reducible and irreducible. The former, with the extra
gluon connecting two incoming (or outgoing) valence quark lines, have
both collinear and soft divergences. The latter, with the extra gluon
connecting the incoming and outgoing quark lines, are lack of collinear
divergences. Therefore, only the reducible radiative corrections are leading
and should be resummed into the Sudakov factor mentioned above.
If the main contributions are due to soft
gluon exchange, the case in which the running coupling constant $\alpha_s$
becomes large, lowest-order PQCD will not be reliable.
With the introduction of the extra $b$ dependence, $\alpha_s$ remains small
as long as $b$ is small. The effect of Sudakov suppression provides the
exact mechanism of confining the scattering process
in the short-distance region. Therefore, the modified PQCD formalism
is relatively self-consistent.
The method to calculate pion Compton
scattering based on these diagrams
is similar to that developed for electromagnetic form factors \cite{LS}.

We start with the factorization formula for $H(s,t)$, keeping
transverse momentua ${\bf k}_T$ carried by  valence quarks in
the pion wave function $\psi$ and the hard-scattering kernel $T_H$,
\begin{eqnarray}
H(s,t)&=&\sum_{l=1}^2\int_0^1\; \d x_{1}\d x_{2}
\d^2{\bf k}_{T_1}\d^2{\bf k}_{T_2}\;
\psi(x_2,{\bf k}_{T_2},p_{2}) \nonumber \\
& &\times
T_{H_l}(x_i,s,t,{\bf k}_{T_i})\,
\psi(x_1,{\bf k}_{T_1},p_{1})\; .
\label{2}
\end{eqnarray}
If ${\bf k}_T$ in $T_H$ is assumed to give higher-power
$({\bf k}_T^2/Q^2)$ correction and thus neglected, the ${\bf k}_T$
integration can be performed, and eq.~(\ref{2}) leads to
the standard factorization formula \cite{BL}.
However,  this approximation
is not proper when the exchanged gluon becomes soft.
The contribution to the hard scattering
from each diagram in fig.~3, obtained by contracting the two photon vertices
with $-g^{\mu\nu}$, is listed in table~1.
All the contributions can be grouped into two terms $(l=1,\;2)$ using the
permutative symmetry.

Rewriting eq.~(\ref{2}) in terms of the Fourier transformed functions,
and inserting
the large-$b$ asymptotic behavior of the wave function \cite{BS}, we have
\begin{eqnarray}
H(s,t)&=& \sum_{l=1}^2 \int_{0}^{1}\d x_{1}\d x_{2}\,
\phi(x_{1})\phi(x_{2})
\int_{0}^{\infty} b\d b
{\tilde T}_{H_l}(x_i,s,t,b)
\nonumber \\
& & \times \exp[-S(x_i,b,Q)]\; .
\label{15}
\end{eqnarray}
where $b$, introduced by the
Fourier transform, is the separation between the two valence quark
lines as stated before.
In the PQCD approach the pions are assumed to be on-shell,
$p_i^2=0$. Hence, we have $t=-2Q^2$, as obtained from eq.~(\ref{q})
by setting $s_i$ to zero.
Note the extra Sudakov factor $\exp(-S)$ compared to the standard
factorization formula, which arises from the all-order
summation of the collinear
enhancements in radiative corrections to fig.~3.
The exponent $S$ is written as \cite{LS}
\begin{equation}
S(x_{1},x_{2},b,Q,w)=\sum_{i=1}^{2}(s(x_{i},b,Q)+s(1-x_{i},b,Q))-
\frac{2}{\beta_{1}}{\rm ln}\frac{\hat{w}}{-\hat{b}}\; ,
\label{16}
\end{equation}
with
\begin{eqnarray}
s(\xi,b,Q)&=&\frac{A^{(1)}}{2\beta_{1}}\hat{q}\ln\left(\frac{\hat{q}}
{-\hat{b}}\right)+
\frac{A^{(2)}}{4\beta_{1}^{2}}\left(\frac{\hat{q}}{-\hat{b}}-1\right)-
\frac{A^{(1)}}{2\beta_{1}}(\hat{q}+\hat{b})
\nonumber \\
& &-\frac{A^{(1)}\beta_{2}}{16\beta_{1}^{3}}\hat{q}\left[\frac{\ln(-2\hat{b})
+1}{-\hat{b}}-\frac{\ln(2\hat{q})+1}{\hat{q}}\right]
\nonumber \\
& &-\left[\frac{A^{(2)}}{4\beta_{1}^{2}}-\frac{A^{(1)}}{4\beta_{1}}
\ln\left(\frac{e^{2\gamma-1}}{2}\right)\right]
\ln\left(\frac{\hat{q}}{-\hat{b}}\right)
\nonumber \\
& &-\frac{A^{(1)}\beta_{2}}{32\beta_{1}^{3}}\left[
\ln^{2}(2\hat{q})-\ln^{2}(-2\hat{b})\right]\; .
\end{eqnarray}
The variables ${\hat q}$, ${\hat b}$ and ${\hat w}$ are defined by
\begin{eqnarray}
{\hat q} &\equiv & {\rm ln}\left [\xi Q/\Lambda\right ]
\nonumber \\
{\hat b} &\equiv & {\rm ln}(b\Lambda)
\nonumber \\
{\hat w} &\equiv & {\rm ln}(w/\Lambda)\; ,
\label{11}
\end{eqnarray}
where the scale parameter $\Lambda\equiv
\Lambda_{\rm QCD}$ will be set to 0.1 GeV.
The coefficients $\beta_i$ and $A^{(i)}$ are
\begin{eqnarray}
& &\beta_{1}=\frac{33-2n_{f}}{12}\;,\;\;\;\;\beta_{2}=\frac{153-19n_{f}}{24}\;
,
\nonumber \\
& &A^{(1)}=\frac{4}{3}\;,
\;\;\;\; A^{(2)}=\frac{67}{9}-\frac{\pi^{2}}{3}-\frac{10}{27}n_
{f}+\frac{8}{3}\beta_{1}\ln\left(\frac{e^{\gamma}}{2}\right)
\label{12}
\end{eqnarray}
with $n_f=3$ the number of quark flavors and $\gamma$ the Euler constant.
The Sudakov factor is always  less than 1 as explained in \cite{LS},
and decreases quickly in the large-$b$ region.
The function $\phi$, obtained by factoring the $Q$ and $b$ dependences
from the transformed wave function
into Sudakov logarithms, is
taken as the Chernyak and Zhitnitsky model \cite{CZ1}
\begin{equation}
\phi^{CZ}(x)=\frac{15f_{\pi}}{\sqrt{2N_{c}}}\,x(1-x)(1-2x)^{2}\; ,
\end{equation}
where $N_{c}=3$ is the number of colors.

The transformed hard scatterings ${\tilde T}_{H_l}$
are given by
\begin{eqnarray}
{\tilde T}_{H_1}
&=&\frac{16\pi{\cal C}_F(e_u^2+e_d^2)\alpha_s(w_1)}{(1-x_1)(1-x_2)}
K_0\left(\sqrt{|r_1|}b\right)\left(\frac{[(1-x_1)t+u][(1-x_2)t+u]}{s^2}
\right.
\nonumber \\
& &\left.+\frac{[(1-x_1)t+s][(1-x_2)t+s]}{u^2}-4(1-x_2)\right)
\label{h1}
\end{eqnarray}
from the classes of fig.~3a-3c, and
\begin{eqnarray}
{\tilde T}_{H_2}
&=&32\pi{\cal C}_F e_u e_d\alpha_s(w_2)\left[\theta(-r_2)
K_0\left(\sqrt{|r_2|}b\right)
+\theta(r_2)\frac{i\pi}{2}H_0^{(1)}\left(\sqrt{r_2}b\right)\right]
\nonumber \\
&  & \times\left(\frac{1}{x_1(1-x_1)}-
\frac{(1+x_2-x_1 x_2)t^2+(1+x_2-x_1)ut}{x_2(1-x_{1})s^2}\right.
\nonumber \\
& &\;\;\;\;\left.+\frac{1}{x_2(1-x_2)}-
\frac{(1+x_1-x_1x_2)t^2+(1+x_1-x_2)st}{x_1(1-x_{2})u^2}\right)
\label{hh}
\end{eqnarray}
from the classes of fig.~3d-3e with
\begin{eqnarray}
r_1=x_1x_2t,\;\;\;\;\;\;r_2=x_{1}x_{2}t+x_1u+x_2s\; .
\end{eqnarray}
$K_{0}$ and $H_0^{(1)}$ in eqs.~(\ref{h1}) and (\ref{hh}) are the
Bessel functions in the standard notation. The imaginary contribution comes
from the case in which the exchanged gluons in fig.~3d and 3e are on-shell,
or $r_2$ vanishes.
The argument $w_l$ of $\alpha_s$
is defined by the largest mass scale in the hard scattering,
\begin{equation}
w_1=\max\left(\sqrt{|r_1|},\frac{1}{b}\right),\;\;\;\;\;\;
w_2=\max\left(\sqrt{|r_2|},\frac{1}{b}\right)\; .
\label{9}
\end{equation}
As long as $b$ is small, soft $r_l$  does not lead to large $\alpha_s$.
Therefore, the non-perturbative region in the modified factorization
is characterized by large $b$, where Sudakov suppression is strong.
Eq.~(\ref{15}), as a perturbative expression, is thus
relatively self-consistent compared to
the standard factorization. Since the singularity associated with
$r_2=0$ is not even suppressed by the pion wave function, Sudakov effects
are more crucial in Compton scattering \cite{MF} than in the case of
form factors. The numerical outcomes of eq.~(\ref{15})
have been obtained and will be shown in section 5.

Following the similar procedures, we derive the first invariant
amplitude $H_1$.
The extraction of $H_1$ can be performed
by contracting the two photon vertices with $e^{(1)\mu}e^{(1)\nu}$.
The derivation of $H_1$ is much simpler than that of $H_2$, because
$e^{(1)}$ is orthogonal to all of the momenta $p_i$ and $q_i$.
The contribution to the hard scattering associated with $H_1$ from each diagram
in fig.~3 is also listed in table.~1.
The modified perturbative expression for $H_1$ is given by
\begin{eqnarray}
H_1(s,t)&=& \sum_{l=1}^2 \int_{0}^{1}\d x_{1}\d x_{2}\,
\phi(x_{1})\phi(x_{2})
\int_{0}^{\infty} b\d b
{\tilde T}^{(1)}_{H_l}(x_i,s,t,b)
\nonumber \\
& & \times \exp[-S(x_i,b,Q)]\;,
\label{ph1}
\end{eqnarray}
where
\begin{eqnarray}
{\tilde T}^{(1)}_{H_1}
=\frac{8\pi{\cal C}_F(e_u^2+e_d^2)\alpha_s(w_1)}{(1-x_1)(1-x_2)}
\left[\frac{u}{s}+\frac{s}{u}+4-2x-2y\right]
K_0\left(\sqrt{|r_1|}b\right)
\label{1h1}
\end{eqnarray}
from the classes of fig.~3a-3c, and
\begin{eqnarray}
{\tilde T}^{(1)}_{H_2}&=&16\pi{\cal C}_Fe_u e_d\alpha_s(w_2)
\left[\theta(-r_2)K_0\left(\sqrt{|r_2|}b\right)
+\theta(r_2)\frac{i\pi}{2}H_0^{(1)}\left(\sqrt{r_2}b\right)\right]
\nonumber\\
& &\times\left[\frac{1}{x_1(1-x_1)}+\frac{1}{x_2(1-x_2)}
-\frac{t}{x_2(1-x_1)s}-\frac{t}{x_1(1-x_2)u}\right]
\label{1hh}
\end{eqnarray}
from the classes of fig.~3d-3e.
The expressions for the Sudakov exponents and for $w_l$ are the same as
before.
\vskip 2.0cm

\section{Numerical Results and
the Cross Section of $\gamma \gamma \to \pi^+ \pi^-$}

Based on the sum rules and the modified perturbative expressions
for $H_1$ and $H_2$ in the previous sections, we compute the
magnitudes of the two helicities. Sum rule predictions are
obtained from eq.~(\ref{h2}) with
the substitution of $s_0=0.6$ and $M^2=4$ GeV$^2$.
The modified PQCD formula for $H_1$ in
eq.~({\ref{ph1}) is evaluated numerically, and $H_2$ is derived by
$H_2=H-H_1$, where the values of $H$ have been given in
{}~\cite{coli1,coli}.
Results of $H_1$ and $H_2$ at different photon scattering angles
$\theta^*$ are shown in
fig. 4a and 4b respectively, in which $|H_i|$ denotes the magnitude of
$H_i$. Note that sum rule predictions for
$H_1$ are negative, and those for $H_2$ are positive.
It is observed that
sum rule results decrease more rapidly with momentum transfer
$|t|$ \cite{BF}, and have weaker angular dependence
compared to PQCD ones.
The PQCD predictions are always larger than those
from sum rules at $\theta^*=50^o$ $(-t/s=0.6)$,
and are always smaller at $\theta^*=15^o$ $(-t/s=0.2)$ in the range
$4 < |t| < 16$ GeV$^2$. It implies that large-angle Compton scattering
might be dominated by perturbative dynamics. These two approaches overlap
at $\theta^*=40^o$ $(-t/s=0.5)$
and at $|t|=4$ GeV$^2$, showing the transition of
pion Compton scattering to PQCD. The transition scale is higher at smaller
angles. Basically, the behaviors of $H_1$ and $H_2$ are similar to that
of $H$ as shown in \cite{coli1}.

With the knowledge of $H_1$ and $H_2$, we compute the cross section of
pion Compton scattering.
The expression for the differential cross section of pion Compton scattering
in the Breit frame is derived in Appendix C:
\begin{equation}
\frac{\d \sigma}{\d\cos\theta^*}=\frac{|H_1|^2+|H_2|^2}{256\pi t}
\left(\frac{s-u}{s}\right)^3\;,
\label{csb}
\end{equation}
which yields the results exhibited in fig. 5.
In the angular range we are investigating, the PQCD and sum rule methods
predict opposite dependence on $\theta^*$: PQCD results increase, while
sum rule results decrease, with the photon scattering angle. The reason
for the difference is that the increase
of the amplitudes with $\theta^*$ is not sufficient to overcome the increase
of the incident flux (see Appendix C). At a fixed angle, the differential
cross section, similar to $H_1$ and $H_2$,
drops as $|t|$ grows. Again, the transition scale
is around 4 GeV$^2$ for $\theta^*=40^o$.

The differential cross section of pion Compton scattering from a polarized
photon has been analyzed based on the standard factorization formula
in \cite{MT}. It is worthwhile to compare their predictions with ours
from the modified perturbative formalism.
Note that in \cite{MT} the coupling
constant $\alpha_s$ is regarded as a
phenomenological parameter and set to 0.3, while
we consider the running of $\alpha_s$ due to
the inclusion of radiative corrections, and its cutoff is determined
by Sudakov suppression. Furthermore, our perturbative calculation
is self-consistent in the sense that short-distance
(small-$b$) contributions dominate.

We concentrate on the two specific processes:
$\gamma_{\rm R}\pi\to\gamma_{\rm R}\pi$ and
$\gamma_{\rm L}\pi\to\gamma_{\rm R}\pi$, where $\gamma_{\rm R}$
($\gamma_{\rm L}$) denotes a photon with right-handed (left-handed)
polarization. In our approach the amplitude of
$\gamma_{\rm R}\pi\to\gamma_{\rm R}\pi$ is derived in Appendix C as
\begin{equation}
{\cal M}_{\rm RR}=\frac{H_1-H_2}{2}\;,
\label{RR}
\end{equation}
and that of
$\gamma_{\rm L}\pi\to\gamma_{\rm R}\pi$ is
\begin{equation}
{\cal M}_{\rm LR}=-\frac{H_1+H_2}{2}\;.
\label{LR}
\end{equation}
The analysis in \cite{MT} was performed in the center-of-mass frame,
in which the differential cross section
is written as
\begin{equation}
\frac{\d \sigma}{\d\cos\theta}=\frac{|{\cal M}|^2}{32\pi s}\;,
\label{csc}
\end{equation}
with $\theta$ the center-of-mass scattering angle,
$\cos\theta=(t-u)/s$.
Substituting eqs.~(\ref{RR}) and (\ref{LR}) into (\ref{csc}), we derive
the modified PQCD predictions, evaluated at $|t|=4$ GeV$^2$,
whose dependence on $\theta$
is shown in fig.~6a, along with the corresponding results
obtained in \cite{MT} using the Chernyak and Zhitnitsky wave function.
It is found that the behavior of the differential cross section
with respect to $|t|$ and $\theta$
is similar to that in the Breit frame.
The phase angles associated with ${\cal M}_{\rm RR}$
and ${\cal M}_{\rm LR}$ are shown in fig.~6b. Good agreement between
the two predictions is observed in both of the figures, which justifies
our perturbative calculation. Note that ${\cal M}_{\rm LR}$ is real in the
standard PQCD approach \cite{MT}, and its phase angle vanishes.

The expressions presented in this paper can be further applied to
another process, two-pion photoproduction $\gamma\gamma\to
\pi^+\pi^-$. By a simple interchange of
$s$ and $t$ in eqs.~(\ref{h2}) and (\ref{ph1}),
we write down the sum rules and the modified perturbative
expressions describing the process at
intermediate Mandelstam variables. The motivation to study this process
is that experimental data \cite{B}
are available for the total invariant mass of
the two pions, $M(\pi^+\pi^-)=\sqrt{s}$, from 0.3 upto 1.5 GeV and for
$|\cos\theta|<0.6$, $\theta$ being the
center-of-mass scattering angle. It is easy to check
that the resulting sum rules for $H_i$ involved in pion photoproduction
are symmetric in $\theta$ and $\pi-\theta$, as expected.

However, restriction exists in
the application due to the strong resonant
contribution of $f_2(1270)$ appearing
above $s=1$ GeV$^2$, which is not considered in our analysis.
The region, in which it is possible to compare our predictions
with data, is then limited to below the $f_2(1270)$ resonance.
At such an energy scale, the PQCD formalism,
even the modified one, can barely be reliable. Therefore, we will
concentrate on the sum rule method.
Another reason for favoring the sum rule formalism is due to
the angular range of the data. It suffices to study the behavior of $H_i$ for
$0 <\cos\theta <0.6$, or, $0.2 < -t/s <0.5$, because of the
angular symmetry mentioned above. In this range sum rule
predictions are found to be dominant
from the study of pion Compton scattering.
On the other hand, the sum rule method does not work at very low energy,
because the OPE of the correlator
is not applicable when relevant scales go below 1 GeV$^2$.
Due to these constraints, we will investigate only the comparision
of sum rule results with the experimental data in the region of
$M(\pi^+\pi^-)$ around 1 GeV and of $|\cos\theta|<0.6$.
An agreement of our
predictions with the data will justify the sum rule formalism given above.

At $s=1$ GeV$^2$ the asymptotic expressions for the
perturbative spectral densities and the gluonic power corrections shown in
section 3 are not appropriate, since $s$, $t$ and the virtualities $s_i$
are of the same order of magnitude.
It is easy to find that every term in the full
series of $\rho_i^{\rm pert}$,
shown in Appendix B, are equally important at this scale.
This fact
requires us to employ the full expressions for $\rho_i^{\rm pert}$ and
$\rho_i^{\rm gluon}$ in the sum rule analysis of the process.
Note that the quark power corrections $C_i^{\rm quark}$ are exact.
However, $\rho_i^{\rm gluon}$ is too complicated to obtain its
complete formula.
The difficulty can be overcomed, if there indeed exists a largest stable
region for $H_i$ at some best $s_0$.
Since the gluonic power correction should be
about of the same order as the quark one,
the best $s_0$, obtained from the stability analysis based only on
$\rho_i^{\rm pert}$ and $C_i^{\rm quark}$, is close to the exact
value. As an approximation, we perform the stability analysis, say,
of $H_2$, by considering $\rho_2^{\rm pert}(s \leftrightarrow t)$
given in Appendix B
and $C_2^{\rm quark}(s \leftrightarrow t)$ at $M(\pi^+\pi^-)=1$ GeV.
The best $s_0$ is found to
be 0.3 GeV$^2$, for which $H_2$ does not vary much as $M^2>4$ GeV$^2$
for different $\theta$. Once the approximate best $s_0$ is determined,
we compute $H_1$ and $H_2$ in the large $M^2$ limit, the region where
power corrections are negligible. By this means, the difficulty from
the gluonic power corrections is avoided,
and we need to evaluate only the perturbative spectral densities.
Following the above procedures,
we obtain $H_1\approx -0.2$ and $H_2\approx 0.1$ for different $\theta$.
We emphasize that at such a low energy scale power corrections must
play an important role, and our results should be regarded as a rough
estimation at most.

Substituting the approximate values of $H_1$ and $H_2$ into
eq.~(\ref{csc}), the total cross section $\sigma(\gamma\gamma\to
\pi^+\pi^-)$ at $M(\pi^+\pi^-)=1$ GeV
for $|\cos\theta|<0.6$ is simply derived as 113 nb.
The cross section at other values of $M(\pi^+\pi^-)$ around 1 GeV
can be computed in a similar way, and is found to be almost constant.
Results are shown in fig.~7, along with part of
the experimental data obtained by the MarkII collobaration \cite{B}.
It is obvious that our sum rule estimation coincides with
the data very well at $M(\pi^+\pi^-)=1$ GeV. Above 1 GeV the data points rise
rapidly due to the $f_2(1270)$ resonance as mentioned before.
Below 1 GeV the data show a slow falloff, and deviate away from our
predictions, indicating that the sum rule method is not applicable
at very low energy.
Our formalism can be easily generalized to study another similar process
$\gamma\gamma\to\pi^0\pi^0$, for which experimental data are also
available \cite{MA}.

\section{Conclusions}

In this work we have extended the sum rule analysis to
pion Compton scattering,
compared their predictions with those from the modified factorization
theorems, and shown that there is a clear overlap between the two approaches.
We have given two individual sum rules and the modified perturbative
expressions for the
invariant helicity amplitudes to lowest order in $\alpha_s$.
A detailed numerical analysis shows that the transition to PQCD in pion
Compton scattering appears at $|t|=4$ GeV$^2$ and at $\theta^*=40^o$ (in
the Breit frame).
This suggests that the sum rule and PQCD methods are complementary tools
in the description of exclusive reactions, and can help
locate their transition region by studying the
power-law falloff of the corresponding amplitudes.
Note that the stability of the sum rules is observed only when the modified
phenomenological model is adopted.
The values of the duality interval and the Borel mass are fixed at
0.6 and 4 GeV$^2$ respectively, values which
are comparable to those commonly used
in the sum rule analysis of pion form factor.

We have also compared the modified PQCD predictions with those obtained by
the standard perturbative calculation, in the center-of-mass frame
\cite{MT}, and good agreement is observed. Note that
our modified formalism does not involve a free parameter, and is
relatively self-consistent. We have also shown, along the way,
how to relate the helicities $H_i$ to the description
of pion Compton scattering from a polarized photon.
As a last step, we have extended the sum rule
formalism to the crossed process $\gamma\gamma\to\pi^+\pi^-$ by
simple crossing of $s$ and $t$, and our predictions match
the data of pion photoproduction.
A more convincing justification of the formalism presented in this paper
can be found from the study of
other similar processes, such
as proton Compton scattering \cite{FZK}.
In our analysis, which is restricted to lowest order in $\alpha_s$,
the sum rules for the helicities are real, and therefore, issues related
to the perturbative and non-perturbative nature of the phases of
Compton scattering cannot be addressed.
We leave the discussion of these issues to future work.
\vskip 0.5cm
\centerline{\bf Acknowledgements}
We are grateful to George Sterman for helpful discussions.
We thank E. Berger, G. Bodwin, L. Bergstrom, Yu.L. Dokshitzer,
T.H. Hansson, R.R. Parwani, H. Rubinstein, M. Sotiropoulos
A.R. White and C. Zachos for illuminating conversations.
C.C. thanks Prof. Bo Andersson and the Theory Department at Lund University,
and the Physics Dept. of Academia Sinica of Taiwan for hospitality.

\renewcommand{\theequation}{A.\arabic{equation}}
\setcounter{equation}{0}
\vskip 1cm \noindent
\noindent {\large\bf Appendix A. Kinematics. }
\vskip 3mm \noindent

This appendix is intended to provide a more detailed information on the
methods employed in the calculation of the spectral densities of Compton
scattering. Due to the great complexity involved in
the analysis, we do not show every step of the computation, but
illustrate the evaluation of some important integrals, especially those
related to the new projection which
lead to the sum rule for the invariant amplitude $H_1$.
Although the projector
is a complex 4-vector, it is possible to prove that all the
complex contributions cancel. The methods given in this appendix
can be applied to a large class
of Compton-type processes. The notations here are
the same as in \cite{co,co1}.



In the Breit frame of the incoming pion we have the Mandelstam invariants
\begin{eqnarray}
& &t=(p_2-p_1)^2=s_1+s_2 -2 Q^2- {s_1 s_2 \over 2 Q^2}\;,
\nonumber \\
& &u=(p_2-q_1)^2=2 Q^2-s +{{s_1 s_2}\over 2 Q^2}\;,
\end{eqnarray}
where $Q^2$ acts as a large parameter in the scattering process:
\beq
\label{qquadro}
Q^2={1\over 4} (s_1+s_2-t + \delta) = {1 \over 4} (s+u+\delta)\;,
\eeq
with
\beq
\delta=\sqrt{(s_1+s_2-t)^2 -4 s_1 s_2} = {4 Q^4 - s_1 s_2\over 2 Q^2}.
\label{deltadef}
\eeq
Note that $t=-2 Q^2$ at $s_1=s_2=0$.

It is convenient to introduce light-cone variables for the momenta.
We define
\beqa
& &q_1 = q_1^+ \bar{v} +q_1^- v +q_{1\perp},   \nonumber\\
& &p_1=Q \bar{v} +{s_1\over 2 Q} v,\quad\quad\quad   p_2={s_2\over 2
Q} \bar{v} + Q v,
\eeqa
where
\beqa
\bar{v}={1\over\sqrt{2}} (1,1,{\bf 0}_\perp), \quad \quad \quad
v={1\over\sqrt{2}} (1,-1,{\bf 0}_\perp)
\eeqa
satisfy the relations $v^2 = \bar{v}^2=0$ and $v\cdot\bar{v}=1$.
In our notation
\beq
p^{\pm}= {1\over \sqrt{2}} (p^0 \pm p^3)\;.
\eeq
The covariant expressions for $q_1^\pm$ can be easily
obtained as
\beqa
& &q_1^+={(s-2 Q^2)(2 Q^2-s_2)\over 2Q\delta},
\nonumber \\
& &q_1^-={(2 Q^2-s_1)(2 Q^2 s-s_1 s_2)\over 4Q^3\delta}\;.
\eeqa
The expressions for $q_2^\pm$ are similar.

In the frame specified above, all the transverse momenta are carried
by the two photons, which are on shell $(q_1^2,\;q_2^2=0)$,
and carry physical polarizations. The polarization vectors are defined by
\beqa
& &e^{(1)\lambda}={N^{\lambda}\over \sqrt{-N^2}},\quad\quad\quad
e^{(2)\lambda}={P^{\lambda}\over \sqrt{-P^2}},
\nonumber \\
& &N^{\lambda}=\epsilon^{\lambda\mu\nu\rho}P_{\mu}r_{\nu}R_{\rho},
\quad\quad\quad
P^{\lambda}=\nu_1 p_1^\lambda+ \nu_2 p_2^\lambda +{\alpha\over2} R^\lambda ,
\nonumber \\
& &\nu_1 = {p_1\cdot p_2 - s_2},\quad\quad\quad
        \nu_2=p_1\cdot p_2-s_1,
\nonumber \\
& &R=q_1+q_2, \quad \quad \quad r=q_2 -q_1\;,
\eeqa
with
\beq
\alpha=(2/t)(\nu_1 p_1\cdot R + \nu_2 p_2\cdot R)\;.
\eeq
They satisfty the normalization conditions
\beqa
e^{(i)} \cdot q_1 &=& e^{(i)}\cdot q_2 = 0, \nonumber \\
e^{(i)}\cdot e^{(j)} &=& -\delta_{ij}\;.
\label{eortho}
\eeqa
These relations hold for all positive $s_1$ and $s_2$, whether
or not they are equal, and for $\nu_1$ and $\nu_2$ chosen as above.

In this frame the projector $n^\mu=(n^+,n^-,n_\perp)$ is given by
\beqa
n^\pm &=& \left(\nu_1 {p_1}^\pm + {\nu_2}{p_2}^\pm +{ \alpha\over 2}({q_1}^\pm
+
{q_2}^\pm) \right)/\sqrt{-P^2}, \nonumber \\
n_\perp &=&{\alpha \over 2\sqrt{-P^2}}(q_1+q_2)_{\perp} + i e^{(1)}\;,
\label{enne}
\eeqa
which is a complex vector.

All the integrals involved in the computation of
the leading spectral functions corresponding to fig.~1a
appear in the forms
\beqa
I[f(k^2,k\cdot p_1,...)]&\equiv& \int
d^4k   f(k^2,k\cdot
p_1,...){\delta_+(k^2)\delta_+((p_1-k)^2)
\delta_+((p_2-k)^2)\over (p_1-k + q_1)^2}\;,
\nonumber \\
I'[f(k^2,k\cdot p_1,...)]&\equiv& \int d^4k
f(k^2,k\cdot p_1,...)\delta_+(k^2)\delta_+((p_1-k)^2)\delta_+((p_2-k)^2)\;,
\nonumber\\
& &
\label{if}
\eeqa
where $f$ is a functin of arbitrary products containing the internal
momentum $k$. The $\delta$-functions come from the cutting rules applied
to the quark lines except the top one.
The components of $k$ are then fixed at the values
\beqa
& &{\hat k}^+={{Q \left(2 {Q^2} {   s_2} - {s_1} {s_2} \right) }\over
   {4 {Q^4} - {s_1} {s_2}}}\;,
\nonumber \\
& &{\hat k}^-={{Q \left(2 {Q^2} {s_1} -{s_1} {s_2} \right) }\over
   {4 {Q^4} - {s_1} {s_2}}}\;,
\eeqa
and ${\hat k}_{\perp}^2=2{\hat k}^+{\hat k}^-$. Hence, the only
nontrivial integral needed to perform is the one associated with
the polar angle $\theta$ of ${\bf k}_{\perp}$, which can be cast
into the general form
\beq
T(n_1,n_2)= \int_{0}^{2 \pi}d\theta  {\sin^{n_1}\theta
\cos^{n_2}\theta \over A + B  \cos\theta}\;.
\eeq
The expressions for $A$ and $B$ depend on the diagrams we are considering.

As an example we consider the following integral appearing
in the evaluation of $\Delta_1^{\rm pert}$:
\beqa
I[(e^{(1)}\cdot k)^2  n\cdot k]&=&
I[(e^{(1)}_\perp\cdot k_\perp)^2(n^{+}k^- + n^-k^+)]
\nonumber \\
&&  + I[(e^{(1)}_\perp\cdot k_\perp)^2 ({\alpha\over
\sqrt{-P^2}}q_\perp\cdot k_\perp +
i  e^{(1)}_\perp \cdot k_\perp)]\nonumber \\
&=&\left ((n^{+}{\hat k}^- + n^- {\hat k}^+)
{\hat k}_\perp^2  T(2,0)
- {\alpha\over \sqrt{-P^2}} |q_\perp |  {B\over
2}{\hat k}_\perp^2  T(1,1)\right)\nonumber \\
& &\times J(Q^2,s_1,s_2)\;,
\label{I1}
\eeqa
where $J(Q^2,s_1,s_2)$ is a jacobean given by
\beq
J(Q^2,s,s_1,s_2)={Q^2\over 4 (4 Q^2 - s_1 s_2)}.
\eeq
As we have already discussed before,
it is possible to show that the most significant contribution
to the spectral functions comes from the region in which $s_1$ and $s_2$
are close $(s_1\approx s_2)$.

Another useful integral for gluonic power correction corresponding
to the diagram with a gluon from the vacuum attached to each of
the lateral quark lines is
\beq
I[f,m_1,m_2]\equiv \int
d^4k   f{\delta_+(k^2)\delta_+((p_1-k)^2-m_1^2)
\delta_+((p_2-k)^2-m_2^2)\over (p_1-k + q_1)^2}\;.
\label{a2}
\eeq
The relevant quantities are then obtained through the derivatives of this
integral with respect to the two
mass parameters $m_1$ and $m_2$ evaluated at $m_1=m_2=0$. Using a
coincise notation we define
\beq
I_D[f]\equiv\frac{\partial}{\partial m_2}
\frac{\partial}{\partial m_1} I[f,m_1,m_2]|_{m_1,m_2=0}\;.
\eeq
Eq.~(\ref{a2}) can be computed in a similar way with the components of
$k$ fixed at
\beqa
& &{\hat k}'^+={{Q \left( 2 {  m_1} {Q^2} - 2 {  m_2} {Q^2} -
       {  m_1} {   s_2} + 2 {Q^2} {   s_2} -
       {   s_1} {   s_2} \right) }\over
   {4 {Q^4} - {   s_1} {   s_2}}},
\nonumber \\
& &{\hat k}'^-={{q \left( 2 {  m_1} {Q^2} - {  m_1} {   s_1} +
       {  m_2} {   s_1} + 2 {Q^2} {   s_1} -
       {   s_1} {   s_2} \right) }\over
   {4 {Q^4} - {   s_1} {   s_2}}}\;.
\eeqa
The angular integrals involved in this integral are more complicated,
because $A$ and $B$ are given by irrational functions
of $Q^2,s_1$ and $s_2$, which, however,
become rational when the masses are set to zero.

Those integrals containing the factor $k\cdot q_1$ in the numerator,
either the type of eq.~(\ref{if}) or (\ref{a2}), can
be simplified by the following replacement:
\beq
I[f,k\cdot q_1]=p_1\cdot q_1 I[f]+{1\over 2}\left(
I[f,(p_1-k)^2] -I'[f]\right)\;.
\label{repl}
\eeq
The expression for $I'[f]$ has a strong similarity to that
appearing in the calculation of form factors. For example, $I'[1]$ is the
basic integral associated with the pion form factor:
\beq
I'[1]=\int d^4k \delta_+(k^2)\delta_+((p_1-k)^2)\delta_+((p_2-k)^2)=
{\pi\over 2 \delta}\;.
\eeq
In the $I'$-type of integrals the transverse momentum
$q_\perp $ of the photon does not play any role as in the form
factor case in the Breit frame.

At last, for illustrative purposes we compute a simple example
of eq.~(\ref{a2}). Consider
\beqa
I[1,m_1,m_2]
&=& \int d^4 k {\delta_+(k^2)\delta_+((p_2-k)^2 - m_2^2)
\delta_+((p_1-k)^2-m_1^2)\over (p_1 - k + q_1)^2}
\nonumber \\
&=&(-2\pi i)^3 {1\over 4 \delta} T(0,0)
\label{disc}
\eeqa
with
\beqa
T(0,0)&=&\int_{0}^{2 \pi}{d\theta  \over A +
 B  \cos\theta} \nonumber \\
&=&{2\pi \over {\left( A^2 - B^2\right)}^{1/2}}\;,
\eeqa
\beqa
A&=& s + m_1 -2 {\hat k}'^+ (p_1+q_1)^- -2 {\hat k}'^-
(p_1+q_1)^+ \;,
\nonumber \\
B&=&\sqrt{ 8 q_1^+ q_1^- (2 {\hat k}'^+ {\hat k}'^- -m_1)}\;,
\nonumber \\
A^2-B^2&=& ( -4 {  m_1} {Q^4} + 2 {  m_2} {Q^2} s +
         4 {Q^4} s + 2 {  m_1} {Q^2} {   s_1} \nonumber \\
&&  -  2 {  m_2} {Q^2} {   s_1} - 2 {Q^2} s {   s_1} -
  2 {  m_1} {Q^2} {   s_2} -
         2 {Q^2} s {   s_2} \nonumber \\
&& -   {  m_1} {   s_1} {   s_2} + s {s_1} {s_2} )^2/
  \left(4 {Q^4} - {s_1} {s_2} \right)^2\;.
\eeqa

The explicit expressions of most of the integrals
are rather cumbersome and can not be listed here. The main feature
of the calculation
is the simplification provided by the use of the Breit frame, in which
all the contributions are polynomials of $Q^2,s_1$ and $s_2$.
The spectral functions are
symmetric in $s_1$ and $s_2$, at fixet $s$ and $t$, as expected
from time reversal invariance. For the sum rule investigation of
$H_i(s,t)$ at moderate energy scales,
an expansion of the numerator of each term in the OPE
up to leading virtualities of $s_1$ and $s_2$ is sufficient.
We have examined this approximation
numerically for different $s$ and $t$, and found that the inclusion
of terms with higher power of $s_1$ and $s_2$ gives only small modification.
Based on the key integrals in eqs.~(\ref{if}) and (\ref{a2}),
it is possible to work out the expressions for the perturbative
spectral functions, and the gluonic and quark power corrections given in
sections 2 and 3. We refer in particular to \cite{co} where the
method of calculation of the power corrections has been developed.

\renewcommand{\theequation}{B.\arabic{equation}}
\setcounter{equation}{0}
\vskip 1cm \noindent
\noindent {\large\bf Appendix B. Perturbative Spectral Densities }
\vskip 3mm \noindent

In this appendix we show the full expressions of the perturbative
spectral densities for the invariant helicity amplitudes involved in
pion Compton scattering. The derivation follows the points
outlined in Appendix A, which is simplified drastically in the Breit
frame of the pion. We summarize the results for
$\rho_1^{\rm pert}$ as
\beqa
&& \rho_1^{\rm pert}={5\over 24 \pi^2}{R_1(Q^2,s,s_1,s_2)\over
Q^4 (2 Q^2 -s)^2(4 Q^4 - s_1 s_2)^5 (2 Q^2 s - s_1 s_2)^2}\;,
\eeqa
where
\beqa
R_1(Q^2,s,s_1,s_2)=\sum_{n=0}^{15}a_{n}(s,s_1,s_2) Q^{2 n}.
\eeqa
with
\beqa
a_0& =& {{{s_1}}^9} {{{s_2}}^9} \nonumber \\
a_1& =& -2 {{{s_1}}^8} {{{s_2}}^8}
   \left( 4 s + {s_1} + {s_2} \right) \nonumber \\
a_2& =& 4 s {{{s_1}}^7} {{{s_2}}^7}
   \left( 7 s + 3 {s_1} + 3 {s_2} \right) \nonumber \\
a_3& =& -4 {{{s_1}}^6} {{{s_2}}^6}
   \left(12 {s^3} +7 {s^2} {s_1} +7 {s^2} {s_2}-8 s {s_1} {s_2} -4
{{{s_1}}^2} {s_2}-4 {s_1} {{{s_2}}^2} \right) \nonumber \\
a_4& =& 16 {{{s_1}}^5} {{{s_2}}^5}
   \left( 3 {s^4} +3 {s^3} {s_1} +3 {s^3} {s_2} -
     16 {s^2} {s_1} {s_2}\right. \nonumber \\
&& \left. -8 s {{{s_1}}^2} {s_2} -
     8 s {s_1} {{{s_2}}^2} -6 {{{s_1}}^2} {{{s_2}}^2} \right)
\nonumber \\
a_5 &=& -16 {{{s_1}}^4} {{{s_2}}^4}
   \left( 13 {s^4} {s_1} +13 {s^4} {s_2} -
     20 {s^3} {s_1} {s_2}-16 {s^2} {{{s_1}}^2}
{s_2} - 16 {s^2} {s_1} {{{s_2}}^2} \right. \nonumber \\
&& \left. -40 s {{{s_1}}^2}{{{s_2}}^2} -
     3 {{{s_1}}^3} {{{s_2}}^2} -3 {{{s_1}}^2} {{{s_2}}^3}
\right) \nonumber \\
a_6 &=& 64 {{{s_1}}^3} {{{s_2}}^3}   \left( 6 {s^5} {s_1} +
6 {s^5} {s_2} + 22 {s^4} {s_1} {s_2} +
     19 {s^3} {{{s_1}}^2} {s_2} + 19 {s^3} {\it s_1} {{{\it s_2}}^2}
\right. \nonumber \\
&& \left. +19 {s^2} {{{s_1}}^2} {{{s_2}}^2} +
     5 s {{{s_1}}^3} {{{s_2}}^2}+ 5 s {{{s_1}}^2} {{{s_2}}^3} +
     2 {{{s_1}}^3} {{{s_2}}^3} \right) \nonumber \\
a_7 &=& -64 {{{s_1}}^2} {{{s_2}}^2}
   \left( 4 {s^6} {s_1} + 4 {s^6} {s_2} + 72 {s^5} {s_1} {s_2}
+ 76 {s^4} {{{s_1}}^2} {s_2} + 76 {s^4} {s_1}
{{{s_2}}^2} + 264 {s^3} {{{s_1}}^2} {{{s_2}}^2} \right. \nonumber \\
&& \left. + 103 {s^2} {{{s_1}}^3} {{{s_2}}^2} +
     103 {s^2} {{{s_1}}^2} {{{s_2}}^3} + 120 s {{{s_1}}^3} {{{s_2}}^3} +
12 {{{s_1}}^4} {{{s_2}}^3} +
     12 {{{s_1}}^3} {{{s_2}}^4} \right) \nonumber \\
a_8 &=& 256 {{{s_1}}^2} {{{s_2}}^2}
   \left( 12 {s^6} + 30 {s^5} {s_1} + 30 {s^5} {s_2} +
     164 {s^4} {s_1} {s_2} + 98 {s^3} {{{s_1}}^2} {s_2}
+  98 {s^3} {\it s_1} {{{\it s_2}}^2} \right.\nonumber \\
 && \left. + 196 {s^2} {{{s_1}}^2} {{{s_2}}^2} +
     42 s {{{s_1}}^3} {{{s_2}}^2}+42 s {{{s_1}}^2}
{{{s_2}}^3} + 23 {{{s_1}}^3} {{{s_2}}^3} \right) \nonumber \\
a_9 &=& -512 {s_1} {s_2} \left( 8 {s^6} {s_1} + 8 {s^6} {s_2} +
     72 {s^5} {s_1} {s_2} + 91 {s^4} {{{s_1}}^2} {s_2}
+ 91 {s^4} {s_1} {{{s_2}}^2} \right.\nonumber \\
&&\left. + 236 {s^3} {{{s_1}}^2} {{{s_2}}^2} +
     92 {s^2} {{{s_1}}^3} {{{s_2}}^2}
+ 92 {s^2} {{{s_1}}^2} {{{s_2}}^3} +
     84 s {{{s_1}}^3} {{{s_2}}^3} + 8 {{{s_1}}^4} {{{s_2}}^3} +
     8 {{{s_1}}^3} {{{s_2}}^4} \right) \nonumber \\
a_{10} &=& 1024 {s_1} {s_2} \left( 12 {s^6} + 30 {s^5} {s_1} +
     30 {s^5} {s_2} + 134 {s^4} {s_1} {s_2}\right. \nonumber \\
&&\left. + 98 {s^3} {{{s_1}}^2} {s_2} +
98 {s^3} {s_1} {{{s_2}}^2} +
 121 {s^2} {{{s_1}}^2} {{{s_2}}^2} +27 s {{{s_1}}^3} {{{s_2}}^2}
+ 27 s {{{s_1}}^2} {{{s_2}}^3} +
2 {{{s_1}}^3} {{{s_2}}^3} \right) \nonumber \\
a_{11} &=& -1024 s \left( 4 {s^5} {s_1} + 4 {s^5} {s_2} +
     72 {s^4} {s_1} {s_2} + 76 {s^3} {{{s_1}}^2} {s_2} +
     76 {s^3} {s_1} {{{s_2}}^2} \right. \nonumber \\
&& \left. + 204 {s^2} {{{s_1}}^2} {{{s_2}}^2} +
     73 s {{{s_1}}^3} {{{s_2}}^2} +
     73 s {{{s_1}}^2} {{{s_2}}^3} +
32 {{{s_1}}^3} {{{s_2}}^3} \right)\nonumber \\
a_{12} &=& 4096 \left( 6 {s^5} {s_1} + 6 {s^5} {s_2} +
     37 {s^4} {s_1} {s_2} + 19 {s^3} {{{s_1}}^2} {s_2} +
     19 {s^3} {s_1} {{{s_2}}^2}\right. \nonumber \\
&&\left. + 32 {s^2} {{{s_1}}^2} {{{s_2}}^2} +
     2 s {{{s_1}}^3} {{{s_2}}^2} +
2 s {{{s_1}}^2} {{{s_2}}^3} +
     2 {{{s_1}}^3} {{{s_2}}^3} \right) \nonumber \\
a_{13} &=& -4096 \left( 13 {s^4} {s_1} + 13 {s^4} {s_2} +
     28 {s^3} {s_1} {s_2} + 8 {s^2} {{{s_1}}^2} {s_2}
+ 8 {s^2} {s_1} {{{s_2}}^2} \right.\nonumber \\
&& \left. + 8 s {{{s_1}}^2} {{{s_2}}^2} +
     {{{s_1}}^3} {{{s_2}}^2} + {{{s_1}}^2} {{{s_2}}^3} \right)
\nonumber \\
a_{14} &=& 16384 s \left( 3 {s^2} {s_1} + 3 {s^2} {s_2} +
     s {s_1} {s_2} + {{{s_1}}^2} {s_2} + {s_1} {{{s_2}}^2} \right)
\nonumber \\
a_{15} &=& -16384 {s^2} \left( {s_1} + {s_2} \right). \nonumber
\eeqa
Obviously, the numerator $R_1$ is a polynomial of the variables
$Q^2$, $s$ and $s_i$.

Similarly, the perturbative spectral density $\rho_2^{\rm pert}$
is given by
\beqa
&& \rho_2^{\rm pert}={5\over 24 \pi^2}{R_2(Q^2,s,s_1,s_2)\over
Q^4 (2 Q^2 -s)^2(4 Q^4 - s_1 s_2)^5 (2 Q^2 s - s_1 s_2)^2}\;,
\eeqa
where
\beqa
&& R_2(Q^2,s,s_1,s_2)=\sum_{n=0}^{15}b_{n}(s,s_1,s_2)Q^{2 n}
\eeqa
with
\beqa
b_{0} &=& -{{{s_1}}^9} {{{s_2}}^9} \nonumber \\
b_{1} &=& 2 {{{s_1}}^8} {{{s_2}}^8} \left( 5 s + {s_1}
 + {s_2} \right) \nonumber \\
b_2 &=& -4 {{{s_1}}^7} {{{s_2}}^7}
   \left( 10 {s^2} + 4 s {s_1} + 4 s {s_2} + {s_1} {s_2}
\right)\nonumber \\
b_3 &=& 4 {{{s_1}}^6} {{{s_2}}^6}
   \left( 20 {s^3} +11 {s^2} {s_1} +11 {s^2} {s_2} -
     6 s {s_1} {s_2} - 2 {{{s_1}}^2} {s_2} -
     2 {s_1} {{{s_2}}^2} \right) \nonumber \\
b_4 &=& -16 {{{s_1}}^5} {{{s_2}}^5}
   \left( 5 {s^4} +9 {s^3} {s_1} +9 {s^3} {s_2}
 - 20 {s^2} {s_1} {s_2} -10 s {{{s_1}}^2} {s_2} -
     10 s {s_1} {{{s_2}}^2} - 8 {{{s_1}}^2} {{{s_2}}^2}
\right) \nonumber \\
b_5 &=& 16 {{{s_1}}^4} {{{s_2}}^4}
   \left( 39 {s^4} {s_1} + 39 {s^4} {s_2}
+ 4 {s^3} {s_1} {s_2} - 2 {s^2} {{{s_1}}^2} {s_2} -
     2 {s^2} {s_1} {{{s_2}}^2}\right. \nonumber \\
& &\left. - 62 s {{{s_1}}^2} {{{\it s_2}}^2}
   -11 {{{s_1}}^3} {{{s_2}}^2} - 11 {{{s_1}}^2} {{{s_2}}^3}
\right) \nonumber \\
b_6 &=& -64 {{{s_1}}^3} {{{s_2}}^3}
   \left( 18 {s^5} {s_1} +18 {s^5} {s_2} +
     86 {s^4} {s_1} {s_2} +65 {s^3} {{{s_1}}^2} {s_2}
+65 {s^3} {s_1} {{{s_2}}^2} \right. \nonumber \\
&& \left. +78 {s^2} {{{s_1}}^2} {{{s_2}}^2} +
     16 s {{{s_1}}^3} {{{s_2}}^2}
+ 16 s {{{s_1}}^2} {{{s_2}}^3} - {{{s_1}}^3} {{{s_2}}^3}
\right) \nonumber \\
b_7 &=& 64 {{{s_1}}^2} {{{s_2}}^2}
   \left( 12 {s^6} {s_1} + 12 {s^6} {s_2} +
     216 {s^5} {s_1} {s_2} + 236 {s^4} {{{s_1}}^2} {s_2}
+236 {s^4} {s_1} {{{s_2}}^2}
\right. \nonumber \\
&&\left.  +760 {s^3} {{{s_1}}^2} {{{s_2}}^2} +
     273 {s^2} {{{s_1}}^3} {{{s_2}}^2}
+ 273 {s^2} {{{s_1}}^2} {{{s_2}}^3} +
     290 s {{{s_1}}^3} {{{s_2}}^3} + 32 {{{s_1}}^4} {{{s_2}}^3} +
     32 {{{s_1}}^3} {{{s_2}}^4} \right) \nonumber \\
b_8 &=& -256 {{{s_1}}^2} {{{s_2}}^2}
   \left( 36 {s^6} + 90 {s^5} {s_1} + 90 {s^5} {s_2}
+464 {s^4} {s_1} {s_2} + 286 {s^3} {{{s_1}}^2} {s_2} +
     286 {s^3} {s_1} {{{s_2}}^2}
\right. \nonumber \\
&& \left. + 516 {s^2} {{{s_1}}^2} {{{s_2}}^2}
+106 s {{{s_1}}^3} {{{s_2}}^2} +
     106 s {{{s_1}}^2} {{{s_2}}^3} + 59 {{{s_1}}^3} {{{s_2}}^3}
     \right)\nonumber \\
b_9& =& 512 {s_1} {s_2} \left( 24 {s^6} {s_1} + 24 {s^6} {s_2} +
     216 {s^5} {s_1} {s_2} + 265 {s^4} {{{s_1}}^2} {s_2}
+265 {s^4} {s_1} {{{s_2}}^2}
\right. \nonumber \\
&&\left. +676 {s^3} {{{s_1}}^2} {{{s_2}}^2} +
     266 {s^2} {{{s_1}}^3} {{{s_2}}^2}
+ 266 {s^2} {{{s_1}}^2} {{{s_2}}^3} +
     231 s {{{s_1}}^3} {{{s_2}}^3} + 20 {{{s_1}}^4} {{{s_2}}^3} +
     20 {{{s_1}}^3} {{{s_2}}^4} \right) \nonumber \\
b_{10}&=& -1024 {s_1} {s_2} \left( 36 {s^6} + 90 {s^5} {s_1} +
     90 {s^5} {s_2} + 414 {s^4} {s_1} {s_2}
+286 {s^3} {{{s_1}}^2} {s_2} + 286 {s^3} {s_1} {{{s_2}}^2}
\right. \nonumber \\
&&\left. + 386 {s^2} {{{s_1}}^2} {{{s_2}}^2} +
     86 s {{{s_1}}^3} {{{s_2}}^2}
+86 s {{{s_1}}^2} {{{s_2}}^3} + 13 {{{s_1}}^3} {{{s_2}}^3}
\right)\nonumber \\
b_{11}&=& 1024 \left( 12 {s^6} {s_1} + 12 {s^6} {s_2} +
     216 {s^5} {s_1} {s_2} + 236 {s^4} {{{s_1}}^2} {s_2} +
     236 {s^4} {s_1} {{{s_2}}^2}\right. \nonumber \\
&&\left. +660 {s^3} {{{s_1}}^2} {{{s_2}}^2} +
     233 {s^2} {{{s_1}}^3} {{{s_2}}^2} +
     233 {s^2} {{{s_1}}^2} {{{s_2}}^3}
+150 s {{{s_1}}^3} {{{s_2}}^3} + 10 {{{s_1}}^4} {{{s_2}}^3} +
     10 {{{s_1}}^3} {{{s_2}}^4} \right) \nonumber \\
b_{12} &=& -4096 \left( 18 {s^5} {s_1} + 18 {s^5} {s_2} +
     111 {s^4} {s_1} {s_2} + 65 {s^3} {{{s_1}}^2} {s_2} +
     65 {s^3} {s_1} {{{s_2}}^2}\right. \nonumber \\
&& \left. +108 {s^2} {{{s_1}}^2} {{{s_2}}^2} +
     12 s {{{s_1}}^3} {{{s_2}}^2} +
     12 s {{{s_1}}^2} {{{s_2}}^3} + 8 {{{s_1}}^3} {{{s_2}}^3}
\right) \nonumber \\
b_{13} &=& 4096 \left( 39 {s^4} {s_1} + 39 {s^4} {s_2} +
     84 {s^3} {s_1} {s_2} + 30 {s^2} {{{s_1}}^2} {s_2}
+30 {s^2} {s_1} {{{s_2}}^2}
\right. \nonumber \\
&& \left. + 22 s {{{s_1}}^2} {{{s_2}}^2} +
     {{{s_1}}^3} {{{s_2}}^2} + {{{s_1}}^2} {{{s_2}}^3} \right)
\nonumber \\
b_{14} &=& -16384 \left( 9 {s^3} {s_1} + 9 {s^3} {s_2} +
     2 {s^2} {s_1} {s_2} + 2 s {{{s_1}}^2} {s_2}
+ 2 s {s_1} {{{s_2}}^2} - {{{s_1}}^2} {{{s_2}}^2} \right)
 \nonumber \\
b_{15} &=& 16384 s \left( 3 s {s_1} + 3 s {s_2} -
2 {s_1} {s_2} \right). \nonumber
\eeqa
Based on the above formulas, it is easy to confirm that those terms
with higher power of $s_1$ and $s_2$ are negligible, and
the asympotic expressions in eq.~(\ref{h2}) are justified.
The full expressions for $\rho_i^{\rm pert}$ listed here are useful in
the analysis of pion photoproduction.

\renewcommand{\theequation}{C.\arabic{equation}}
\setcounter{equation}{0}
\vskip 1cm \noindent
\noindent {\large\bf Appendix C. The Scattering Amplitude and Cross Section}
\vskip 3mm \noindent

In this appendix we derive the expressions for the amplitudes
of pion Compton scattering from a polarized photon
and the cross section appearing in section 5. The differential cross section
of Compton scattering is defined by
\begin{equation}
\d\sigma=\frac{|{\cal M}|^2}{F}\d Q\;,
\end{equation}
where ${\cal M}$ is the scattering amplitude, $F$ is the incident flux,
and $\d Q$ is the phase space of the final states,
\begin{equation}
\d Q=(2\pi)^4\delta^{(4)}(p_1+q_1-p_2-q_2)
\frac{\d {\rm p}_2}{(2\pi)^3 2E_{p_2}}
\frac{\d {\rm q}_2}{(2\pi)^3 2E_{q_2}}\;.
\label{phs}
\end{equation}
$F$ is defined by
\begin{equation}
F=|{\bf v}_{p_1}-{\bf v}_{q_1}|2E_{p_1}2E_{q_1}=4p\cdot q=2s\;,
\label{inf}
\end{equation}
with ${\bf v}_{p}={\bf p}/E_p$ the velocity of the incoming particle.
It is easy to observe that $F$ increases with the photon scattering angle.
Combining eqs.~(\ref{phs}) and (\ref{inf}) we obtain the general expression
\begin{equation}
\d\sigma=\frac{|{\cal M}|^2}{32\pi s}\d \cos\theta
\label{csc1}
\end{equation}
with $\theta$ the center-of-mass scattering angle.
The above formula is exactly eq.~(\ref{csc}) in section 5, which
is employed in the comparision of our predictions with those
from ref.~\cite{MT} and with experimental data of
$\gamma\gamma\to\pi^+\pi^-$ \cite{B}.

Substituting $\cos\theta=(t-u)/s$ into eq.~(\ref{csc1}), we have the
expression which is invariant in both of the center-of-mass and Breit frames.
Then eq.~(\ref{csc1}) can be easily converted into the one
in the Breit frame
using the relation $\sin\theta^*/2=-t/(s-u)$ with $\theta^*$
as defined before. We have
\begin{equation}
\frac{\d \sigma}{\d\cos\theta^*}=\frac{|{\cal M}|^2}{128\pi t}
\left(\frac{s-u}{s}\right)^3\;,
\end{equation}
where the scattering amplitude ${\cal M}$ is given by
\begin{equation}
{\cal M}=M_{\mu\nu}\epsilon_{1T}^{\mu}\epsilon_{2T}^{*\nu}
\end{equation}
with $M_{\mu\nu}$ defined by eq.~(\ref{h1h2}), and $\epsilon_T$
the polarization vector of the photon in the state $T$.
Inserting $|{\cal M}|^2=(|H_1|^2+|H_2|^2)/2$ for
pion Compton scattering from an unpolarized photon into the above
formula, we obtain eq.~(\ref{csb}).

In the case involving a polarized photon, the scattering amplitude
is given by
\begin{equation}
{\cal M}_{\rm RR}=M_{\mu\nu}\epsilon_{1R}^{\mu}\epsilon_{2R}^{*\nu}
\end{equation}
for the process
$\gamma_{\rm R}\pi\to\gamma_{\rm R}\pi$, and
\begin{equation}
{\cal M}_{\rm LR}=M_{\mu\nu}\epsilon_{1L}^{\mu}\epsilon_{2R}^{*\nu}
\end{equation}
for $\gamma_{\rm L}\pi\to\gamma_{\rm R}\pi$.
To compute ${\cal M}_{\rm RR}$ and ${\cal M}_{\rm LR}$ explicitly,
we assign the following momenta in the center-of-mass frame \cite{MT}:
\begin{eqnarray}
& &p_1=P(1,0,0,-1)\;,\quad\quad\quad q_1=P(1,0,0,1)\;,\nonumber\\
& &p_2=P(1,-\sin\theta,0,-\cos\theta)\;,\quad\quad\quad
q_2=P(1,\sin\theta,0,\cos\theta)\;,\nonumber\\
& &\epsilon_{1R,L}=\frac{1}{\sqrt{2}}(0,-1,\pm i,0)\;,\quad\quad\quad
\epsilon_{2R,L}=\frac{1}{\sqrt{2}}(0,-\cos\theta,\pm
i,\sin\theta)\;.
\end{eqnarray}
A straightforward calculation leads to the relations
\begin{eqnarray}
& &e^{(1)}\cdot \epsilon_{1R}=
-e^{(1)}\cdot \epsilon_{1L}=
-e^{(1)}\cdot \epsilon_{2R}^{*}=\frac{i}{\sqrt{2}}\;,\nonumber\\
& &e^{(2)}\cdot \epsilon_{1R}=
e^{(2)}\cdot \epsilon_{1L}=
-e^{(2)}\cdot \epsilon_{2R}^{*}=-\frac{1}{\sqrt{2}}\;,
\end{eqnarray}
based on which
it is then a simple matter to work out eqs.~(\ref{RR}) and (\ref{LR}).

\newpage
Table 1. The expressions of the hard scatterings $T_H$ and $T_H^{(1)}$
corresponding to the diagrams in fig.~3. Here we define
\begin{eqnarray}
& &D_1=x_1x_2t-({\bf k}_{T_1}-{\bf k}_{T_2})^{2}\nonumber \\
& &D_2=x_{1}x_{2}t+x_1u+x_2s-({\bf k}_{T_1}-{\bf k}_{T_2})^{2}
\nonumber
\end{eqnarray}

\[ \begin{array}{lcc}   \hline\hline\\
{\rm Diagram}& T_H/(16\pi\alpha_{s}{\cal C}_F) & T_H^{(1)}/(8\pi\alpha_{s}
{\cal C}_F) \\
\hline   \\
({\rm a})     &{\displaystyle
\frac{-e_{u}^2[(1-x_1)t+u][(1-x_2)t+u]}
{(1-x_1)(1-x_2)s^2D_1}}  &{\displaystyle
\frac{-e_{u}^2u}
{(1-x_1)(1-x_2)sD_1} }\\
        &  &   \\
({\rm b})   &  {\displaystyle
\frac{e_{u}^2}
{(1-x_{1})D_1}} & {\displaystyle
\frac{-e_{u}^2}
{(1-x_{1})D_1}}  \\
        &  &  \\
({\rm c})   &  {\displaystyle
\frac{e_{u}^2}
{(1-x_{1})D_1}} & {\displaystyle
\frac{-e_{u}^2}
{(1-x_{1})D_1}}  \\
        &   & \\
({\rm d})   &  {\displaystyle
\frac{-e_{u}e_d}
{x_1(1-x_{1})D_2}}
&   {\displaystyle
\frac{-e_{u}e_d}
{x_1(1-x_{1})D_2}} \\
        &  &  \\
({\rm e})   & {\displaystyle
\frac{e_{u}e_d[(1+x_2-x_1x_2)t^2+(1+x_2-x_1)ut]}
{x_2(1-x_{1})s^2D_2}}
 & {\displaystyle
\frac{e_{u}e_dt}
{x_2(1-x_{1})sD_2} }
\vspace{0.2cm}\\
\hline\hline
\end{array}   \]

\newpage
\centerline{\large \bf Figure Captions}
\vskip 0.5cm
\noindent
{\bf Fig. 1} Lowest-order diagrams for pion Compton scattering in
the sum rule approach.
\vskip 0.3cm
\noindent
{\bf Fig. 2} Variation of (a) $H_1$ and (b) $H_2$ with $M^2$ for (1)
$s_0=0.7$, (2) $s_0=0.6$, and (3) $s_0=0.5$ GeV$^2$.
\vskip 0.3cm
\noindent
{\bf Fig. 3} Lowest-order diagrams for pion Compton scattering in
the perturbative QCD approach.
\vskip 0.3cm
\noindent
{\bf Fig. 4} Dependence of (a) $|t||H_1|$ and (b) $|t||H_2|$ on $|t|$
derived from the modified PQCD (real lines) and from
QCD sum rules (dashed lines)
for (1) $-t/s=0.6$ ($\theta^*=50^o$) (2) $-t/s=0.5$ ($\theta^*=40^o$),
and (3) $-t/s=0.2$ ($\theta^*=15^o$).
\vskip 0.3cm
\noindent
{\bf Fig. 5} Dependence of $d\sigma/d\cos\theta^*$ on $|t|$
derived from the modified PQCD (real lines) and from
QCD sum rules (dashed lines)
for (1) $-t/s=0.6$ (2) $-t/s=0.5$, and (3) $-t/s=0.2$. Note that
the curve (1) from sum rules is shown by a long dashed line.
\vskip 0.3cm
\noindent
{\bf Fig. 6} Dependence of (a) $S^3\d\sigma/d\cos\theta$,
and of (b) the corresponding phase in degrees on $\cos\theta$
from the modified PQCD (real lines)
and from ref.~\cite{MT} (dashed lines)
for (1) $\gamma_{\rm R}\pi\to\gamma_{\rm R}\pi$
and (2) $\gamma_{\rm L}\pi\to\gamma_{\rm R}\pi$.
Note that $S=\sin(\theta/2)$ here.
\vskip 0.3cm
\noindent
{\bf Fig. 7} Dependence of $\sigma(\gamma\gamma\to\pi^+\pi^-)$ on
$M(\pi^+\pi^-)$ derived from QCD sum rules. Part of experimental
data adopted from \cite{B} are also shown.

\end{document}